\documentstyle[aps]{revtex}
\draft

\begin{document}
\title{Note on Separability of the Werner states in
arbitrary dimensions \footnote[1] { This work was supported in
part by the National Security Agency.}}
\author{Arthur O. Pittenger \dag \footnote[3]{Present address: The Centre
for Quantum Computation, Clarendon Laboratory,
 Oxford University}  and Morton H. Rubin \ddag}
\address{\dag Department of Mathematics and Statistics,
University of Maryland, Baltimore County,
Baltimore, MD 21228-5398}
\address{\ddag Department of Physics,
University of Maryland, Baltimore County,
Baltimore, MD 21228-5398}
\date{February 22, 2000}
\maketitle

\begin{abstract}
 Great progress has been made recently in establishing conditions
for separability of a particular class of Werner densities on the tensor
product space of $n$ $d$--level systems (qudits). In this brief note we
complete the process of establishing necessary and sufficient
conditions for separability of these Werner densities by proving the
sufficient condition for general $n$ and $d$.

\end{abstract}


\bigskip

We consider the $d^{n}$-dimensional Hilbert space
\[
H^{\left[ d^{n}\right] }=H^{\left[ d\right] }\otimes \cdots \otimes H^{\left[
d\right] }
\]
composed of the direct product of $n$ $d$--dimensional Hilbert spaces. As in
\cite{pitrub2} we let $\tilde{j}$ denote the $n$-tuple $j\ldots j$ and
define $\left| \tilde{j}\right\rangle =\left| j\right\rangle \otimes \cdots
\otimes \left| j\right\rangle $. The particular class of generalized Werner
state, $W^{\left[
d^{n}\right] }\left( s\right) ,$ considered here is defined as the convex
combination of the
completely random state $\frac{1}{d^{n}}I^{(n)}$ and an entangled
pure state $\left| \Psi \right\rangle \left\langle \Psi \right| $,
\begin{equation}
W^{\left[ d^{n}\right] }\left( s\right) =\left( 1-s\right) \frac{1}{d^{n}}%
I^{(n)}+s\left| \Psi \right\rangle \left\langle \Psi \right|   \label{werdef}
\end{equation}
where $I^{(n)}$ is the identity operator on the $d^{n}$--dimensional Hilbert
space. To be specific, take
\[
\left| \Psi \right\rangle =\frac{1}{\sqrt{d}}\sum_{j=0}^{d-1}\left| \tilde{j}%
\right\rangle .
\]
The Werner state was originally defined in \cite{Werner} for two qubits. Its
generalization has been applied to study Bell's inequalities and local
reality, and it has served as a test case of separability arguments in a
number of studies. The problem treated here is to determine necessary and
sufficient conditions on the parameter $s$ so that $W^{\left[ d^{n}\right]
}\left( s\right) $ is fully separable. That is
\begin{equation}
W^{[d^{n}]}(s)=\sum_{a}p\left( a\right) \rho ^{\left( 1\right) }\left(
a\right) \otimes \cdots \otimes \rho ^{\left( n\right) }\left( a\right) ,
\label{totsep}
\end{equation}
where the $\rho ^{\left( r\right) }\left( a\right) $ are density matrices on
the respective $d$-dimensional Hilbert spaces and the set of numbers \{$p(a)$%
\} is a probablility distributuion. For references to many of the related
studies and for the relevance of the Werner states to the study of
entanglement the reader can consult \cite{pitrub2,caves,pitrub1,braunstein}.

As shown in \cite{pitrub1}, a necessary condition for separability for all $%
d $ and $n$ follows from the Peres partial transpose condition \cite{peres}
or from a weaker condition that can be proved via the Cauchy-Schwarz
inequality. Specifically suppose $j=j_{1}\ldots j_{n}$ and $k=k_{1}\ldots
k_{n}$ differ in each component: $j_{r}\neq k_{r}$. Let $u$ and $v$ be
indices with $u_{r}\neq v_{r}$ and $\left\{ u_{r},v_{r}\right\} =$ $\left\{
j_{r},k_{r}\right\} .$ Then for fully separable\ states $\rho $
\begin{equation}
\left( \sqrt{\rho _{j,j}}\sqrt{\rho _{k,k}}\right) \geq \left| \rho
_{u,v}\right| ,  \label{necessary}
\end{equation}
where $\rho $ is written as a matrix in the computational basis defined by
the tensor products of $\left| j_{i}\right\rangle \left\langle k_{i}\right|
,1\leq i\leq n$. Choosing $j$ and $k$ appropriately in (\ref{necessary}), we
obtain the necessary condition
\begin{equation}
s\leq \left( 1+d^{n-1}\right) ^{-1},  \label{sepnec}
\end{equation}
and special cases of this condition were derived in \cite
{braunstein,caves0,dur1}, for example.

The remaining challenge has been to show that this necessary condition is
also sufficient, and various partial results have been obtained in the
papers just cited. In particular, (\ref{sepnec}) was shown in \cite{caves}
to be sufficient for all $d$ and $n=2$ and in \cite{pitrub2} sufficiency was
established for $d$ prime and all $n$. In this note we complete the study of
this aspect of the Werner states by proving the sufficiency part of the
following result.\smallskip

{\bf Theorem}: The Werner density $W^{\left[ d^{n}\right] }\left( s\right) $
is fully separable if and only if $s\leq \left( 1+d^{n-1}\right) ^{-1}$%
.\smallskip

The relevant technique combines a representation of fully separable states
presented in \cite{caves} with an induction argument presented in \cite
{pitrub2}. Let $s=\left( 1+d^{n-1}\right) ^{-1}$. Then it is easy to show
that
\begin{equation}
W^{\left[ d^{n}\right] }\left( s\right) =\frac{1}{1+d^{n-1}}\left( \frac{1}{d%
}\sum_{j=0}^{d-1}\left| \tilde{j}\right\rangle \left\langle \tilde{j}\right|
\right) +\frac{d^{n-1}}{1+d^{n-1}}\left( \frac{1}{d^{n}}\left(
I^{(n)}+\sum_{j\neq k}\left| \tilde{j}\right\rangle \left\langle \tilde{k}%
\right| \right) \right) .  \label{wers0}
\end{equation}
Since the first term in (\ref{wers0}) is a sum of separable projections, the
proof reduces to showing the separability of the second term. It is
convenient in what follows to intorduce a set of fixed phases and to show
that
\begin{equation}
\rho ^{(n)}=\frac{1}{d^{n}}\left( I^{(n)}+\sum_{j\neq k}\zeta _{j}\left|
\tilde{j}\right\rangle \left\langle \tilde{k}\right| \zeta _{k}^{\ast
}\right)   \label{rhon}
\end{equation}
where $\{|\zeta _{r}|=1,\;r=0,\cdots ,d-1\}$ is separable.

We proceed by induction. When $n=1$, (\ref{rhon}) becomes
\begin{equation}
\rho ^{(1)}=\frac{1}{d}\left( \sum_{j=0}^{d-1}\sum_{k=0}^{d-1}\zeta
_{j}\left| j\rangle \langle k\right| \zeta _{k}^{\ast }\right)   \label{rho1}
\end{equation}
which is obviously a projection for any choice of the parameters $\zeta _{r}$%
. Now assume that the density matrix of the form in (\ref{rhon}) is fully
separable for $n$; then we shall show that it is fully separable for $n+1$.
Following the ideas in \cite{caves}, for a fixed choice of parameters $\zeta
_{r}$ define
\begin{equation}
{\bf w}^{(m)}=(\zeta _{0}z_{0}^{(m)},\cdots ,\zeta _{d-1}z_{d-1}^{(m)})
\label{wvector}
\end{equation}
where $z_{j}^{(m)}\in \left\{ \pm 1,\pm i\right\} $ and $\zeta _{r}$ is
independent of $m$. We have a total of $4^{d}$ different vectors. It is easy
to check that if we sum over all $m$,
\[
\sum_{m}{\bf w}_{r}^{(m)}=\zeta _{r}\sum_{m}z_{r}^{(m)}=0,\quad
\sum_{m}\left( {\bf w}_{r}^{(m)}\right) ^{2}=\zeta _{r}^{2}\sum_{m}\left(
z_{r}^{(m)}\right) ^{2}=0\text{, and\quad }\sum_{m}\left| {\bf w}%
_{r}^{(m)}\right| ^{2}=4^{d}.
\]

For each ${\bf w}^{(m)}$ define the product state $\rho (m)=\rho ^{\left(
n\right) }\left( {\bf w}^{(m)}\right) \otimes \rho ^{\left( 1\right) }\left(
{\bf z}^{(m)\ast }\right) $ where ${\bf z}^{(m)}$ is equal to ${\bf w}^{(m)}$
with all the $\zeta _{r}$'s equal to $1$ and
\begin{eqnarray*}
\rho ^{\left( n\right) }\left( {\bf w}^{(m)}\right)  &=&\frac{1}{d^{n}}%
\left( I^{(n)}+\sum_{j\neq k}\zeta _{j}z_{j}^{(m)}|\tilde{j}%
\rangle \langle \tilde{k}|\zeta _{k}^{\ast }z_{k}^{(m)\ast }\right)  \\
\rho ^{\left( 1\right) }\left( {\bf z}^{(m)}\right)  &=&\frac{1}{d}\left(
I^{\left( 1\right) }+\sum_{r\neq s}z_{r}^{(m)}\left|
r\right\rangle \left\langle s\right| z_{s}^{(m)\ast }\right) .
\end{eqnarray*}
The state $\rho (m)$ is separable by the induction hypothesis,and it
follows that the convex combination $\rho ^{(n+1)}=\frac{1}{4^{d}}%
\sum_{m}\rho (m)$ is also separable. Now we multiply out the terms:
\begin{eqnarray*}
\rho ^{(n+1)} &=&\frac{1}{d^{n+1}}\left( I^{(n+1)}+I^{(n)}\otimes
A^{(1)}+A^{(n)}\otimes I^{(1)}+B\right)  \\
A^{(1)} &=&\sum_{r\neq s}\left| r\right\rangle \left\langle
s\right| \frac{1}{4^{d}}\sum_{m}z_{r}^{(m)\ast }z_{s}^{(m)} \\
A^{(n)} &=&\sum_{j\neq k}\zeta _{j}\zeta _{k}^{\ast }\left|
\tilde{j}\rangle \langle \tilde{k}\right| \frac{1}{4^{d}}%
\sum_{m}z_{j}^{(m)}z_{k}^{(m)\ast } \\
B &=&\sum_{j\neq k}\sum_{r\neq s}\zeta _{j}\zeta _{k}^{\ast }|%
\tilde{j}\rangle \langle \tilde{k}|\otimes \left| r\rangle \langle s\right|
\frac{1}{4^{d}}\sum_{m}z_{j}^{(m)}z_{k}^{m\ast }z_{r}^{(m)\ast }z_{s}^{(m)}.
\end{eqnarray*}
Since the components of ${\bf w}^{(m)}$ are chosen independently of one
another, $\sum_{m}z_{r}^{(m)\ast }z_{s}^{(m)}=0$ for $r\neq s$;
consequently, $A^{(n)}$ and $A^{(1)}$ vanish. As noted in \cite{caves} the
choice of the ${\bf w}^{(m)}$ also simplifies the remaining term, since for $%
r\neq s$ and $j\neq k$
\[
\frac{1}{4^{d}}\sum_{m}z_{j}^{(m)}z_{k}^{(m)\ast }z_{r}^{(m)\ast
}z_{s}^{(m)}=\delta \left( j,r\right) \delta \left( k,s\right) ,
\]
where $\delta (r,s)$ is the Kronecker delta. Then
\[
\rho ^{\left( n+1\right) }=\frac{1}{d^{n+1}}\left( I^{\left( n+1\right)
}+\sum_{j\neq k}\zeta _{j}\left| \tilde{j}j\rangle \langle \tilde{k}k\right|
\zeta _{k}^{\ast }\right) ,
\]
which is of the same form as (\ref{rhon}) with $n\rightarrow n+1,$ completing
the induction step and the proof of the theorem.
.

\end{document}